\documentclass[a4paper]{jpconf}

\usepackage{epsfig}
\usepackage{amsmath}
\usepackage{amssymb}
\usepackage{mathrsfs}
\usepackage{graphicx}
\usepackage{relsize}
\usepackage{mathtools}
\usepackage{multirow}
\usepackage[hyphens]{url}
\usepackage[breaklinks,colorlinks=true,citecolor=black,linkcolor=black]{hyperref}
\hypersetup{
     colorlinks   = true,
     citecolor    = blue
}

\newcommand{\apj}{ApJ} % The Astrophysical Journal
\newcommand{\apjl}{ApJ} % The Astrophysical Journal Letters
\newcommand{\apjs}{ApJ} % The Astrophysical Journal Supplement
\newcommand{\aj}{AJ} % The Astronomical Journal
\newcommand{\aap}{A\&A} % Astronomy & Astrophysics
\newcommand{\araa}{ARAA} % Annual Review of Astronomy and Astrophysics
\newcommand{\mnras}{MNRAS} % Monthly Notices of the Royal Astronomical Society
\newcommand{\alphavir}{\alpha_\mathrm{vir}}

\newcommand{\mach}{\mathcal{M}}
\newcommand{\macha}{\mathcal{M}_\mathrm{A}}

\newcommand{\kinj}{k_\mathrm{inj}}

\newcommand{\sfr}{\mathrm{SFR}}
\newcommand{\sfrff}{\mathrm{SFR}_\mathrm{ff}}

\newcommand{\sfe}{\mathrm{SFE}}
\newcommand{\msol}{\mbox{$M_{\odot}$}}

\newcommand{\tff}{t_\mathrm{ff}}

\newcommand{\cs}{c_\mathrm{s}}

\newcommand{\km}{\mathrm{km}}
\newcommand{\pc}{\mathrm{pc}}
\newcommand{\AU}{\mbox{AU}}
\newcommand{\s}{\mathrm{s}}
\newcommand{\yr}{\mathrm{yr}}
\newcommand{\Myr}{\mathrm{Myr}}
\newcommand{\Gauss}{\mathrm{G}}
\newcommand{\cm}{\mbox{cm}}
\newcommand{\g}{\mbox{g}}
\newcommand{\G}{\mbox{G}}

\begin{document}
\title{The role of turbulence, magnetic fields and feedback for star formation}

\author{Christoph Federrath}

\address{Research School of Astronomy and Astrophysics, Australian National University, Canberra, ACT~2611, Australia}

\ead{christoph.federrath@anu.edu.au}

\begin{abstract}
Star formation is inefficient. Only a few percent of the available gas in molecular clouds forms stars, leading to the observed low star formation rate (SFR). The same holds when averaged over many molecular clouds, such that the SFR of whole galaxies is again surprisingly low. Indeed, considering the low temperatures, molecular clouds should be highly gravitationally unstable and collapse on their global mean freefall timescale. And yet, they are observed to live about 10--100 times longer, i.e., the SFR per freefall time ($\sfrff$) is only a few percent. Thus, other physical mechanisms must provide support against quick global collapse. Magnetic fields, turbulence and stellar feedback have been proposed as stabilising agents controlling star formation, but it is still unclear which of these processes is the most important and what their relative contributions are. Here I present high-resolution, adaptive-mesh-refinement simulations of star cluster formation that include turbulence, magnetic fields, and protostellar jet/outflow feedback. These simulations produce nearly realistic star formation rates consistent with observations, but only if turbulence, magnetic fields and feedback are included simultaneously.
\end{abstract}

Star formation is inefficient. For example, our entire home galaxy---the Milky Way---only produces about $2\,\msol\,\yr^{-1}$ \cite{ChomiukPovich2011}, while there is plenty of gas available to form hundreds of stars per year \cite{ZuckermanPalmer1974,ZuckermanEvans1974}. Based on the range of typical molecular cloud masses of \mbox{$\sim10^2$--$10^7\,\msol$} and sizes of \mbox{$\sim1$--$100\,\pc$}, the average cloud mass density $\rho$ is a few times \mbox{$10^{-21}$--$10^{-19}\,\g\,\cm^{-3}$}, corresponding to an average particle number density of about \mbox{$10^3$--$10^5\,\cm^{-3}$} for standard molecular composition. This would render the clouds highly unstable and immediately leads to a very short freefall time, \mbox{$\tff=\sqrt{3\pi/(32G\rho)}=0.1$--$1\,\Myr$}, which is at least one order of magnitude shorter than the typical lifetime of molecular clouds. Thus, other mechanisms than thermal pressure must be at work to prevent molecular clouds from collapsing globally and from forming stars at a \mbox{$10$--$100$} times higher rate than observed.

For a long time, the standard picture of star formation was that magnetic fields provide the primary source of stabilizing pressure and tension, which acts against the quick global collapse. Only after the neutral species had slowly diffused through the charged particles over an `ambipolar-diffusion' time-scale of about $10\,\Myr$, would star formation proceed in the central regions of magnetized clouds \cite{MestelSpitzer1956,Mouschovias1976,Shu1983}. This standard theory requires that clouds start their lives dominated by a strong magnetic field, rendering them initially subcritical to collapse. After about one ambipolar diffusion time-scale, magnetic flux was left behind in the cloud envelope, while the mass has increased in the cloud core such that some stars could form in centre. Thus, star formation regulated by ambipolar diffusion predicts a higher mass-to-flux ratio in the cores than in the envelopes of the clouds, which is---however---typically not observed \cite{CrutcherHakobianTroland2009,MouschoviasTassis2009,LunttilaEtAl2009,SantosLimaEtAl2010,LazarianEsquivelCrutcher2012,BertramEtAl2012}.

An alternative scenario is that clouds are in fact collapsing globally, thus initially forming stars at a very high rate, but that the ionization feedback from massive stars eventually terminates the global collapse and disperses the clouds \cite{VazquezSemadeni2015}.

A third alternative is that the observed supersonic random motions \cite{ZuckermanPalmer1974,ZuckermanEvans1974,Larson1981,SolomonEtAl1987,FalgaronePugetPerault1992,OssenkopfMacLow2002,HeyerBrunt2004,SchneiderEtAl2011,RomanDuvalEtAl2011} regulate star formation. In this modern picture, turbulence plays a crucial dual role. On the one hand, the turbulent kinetic energy stabilizes the clouds on large scales and prevents global collapse, on the other hand, it induces local compressions in shocks, because the turbulence is supersonic \cite{MacLowKlessen2004,ElmegreenScalo2004,McKeeOstriker2007}. This generates the initial conditions for star formation, because the local compressions typically produce filaments and dense cores at the intersections of filaments \cite{SchneiderEtAl2012}.

So which scenario is right or wrong? As it turns out, neither of the three is entirely wrong, nor do they explain everything by themselves. In the most recent years, we have come to a deeper understanding of the relative importance of magnetic fields and turbulence through numerical simulations. While the role of turbulence cannot be denied, because it naturally explains most of the observed velocity and density structure of molecular clouds, the role of magnetic fields remained less clear. Previous simulation work suggested that magnetic fields can suppress fragmentation on small scales and hence influence the shape of the initial mass function of stars \cite{PriceBate2007,HennebelleTeyssier2008,PetersEtAl2011} and that the SFR can be reduced by a factor of \mbox{2--3} compared to the non-magnetized case \cite{WangEtAl2010,PadoanNordlund2011,FederrathKlessen2012,MyersEtAl2014}.\footnote{See also earlier work on the effects of magnetic fields on the star formation efficiency \cite{VazquezSemadeniEtAl2005,NakamuraLi2005}.} Importantly however, magnetic fields also launch fast, powerful, mass-loaded jets and outflows from the protostellar accretion disc. These jets and outflows can drive turbulence and alter the cloud structure and dynamics so profoundly that the SFR and the initial mass function might be even more affected by this jet/outflow feedback mechanism \cite{KrumholzEtAl2014,FederrathEtAl2014}.

The aim of this study is to determine the physical processes that make star formation inefficient. We measure which physical mechanisms are relevant for bringing the SFR in agreement with observations and we determine their relative importance, i.e., by what amount they reduce the SFR individually and when acting all in concert.

Section~\ref{sec:methods} summarizes our numerical methods and simulations from which we measure the SFR. Our results are presented in Section~\ref{sec:results}, where we find that purely self-gravitating molecular clouds are indeed highly unstable, while the step-by-step inclusion of turbulence, magnetic fields, and jet/outflow feedback brings the SFR down by one to two orders of magnitude when all of these physical mechanisms act together. We summarize our findings and conclusions in Section~\ref{sec:conclusions}. Some of the results of this work are published in Federrath~(2015) \cite{Federrath2015}, and a featured article about this work with comments and animations of the simulations presented here is posted on \url{http://astrobites.org/2015/04/28/why-is-star-formation-so-inefficient/}.

% SECTION: Numerical simulations
\section{Numerical simulation techniques} \label{sec:methods}

We use the multi-physics, adaptive mesh refinement (AMR) \cite{BergerColella1989} code \textsc{flash} \cite{FryxellEtAl2000,DubeyEtAl2008} in its latest version~(v4), to solve the compressible magnetohydrodynamical (MHD) equations on three-dimensional (3D) periodic grids of fixed side length $L$, including turbulence, magnetic fields, self-gravity and outflow feedback. The positive-definite HLL5R Riemann solver \cite{WaaganFederrathKlingenberg2011} is used to guarantee stability and accuracy of the numerical solution of the MHD equations.

\subsection{Turbulence driving} \label{sec:turbdriving}
Turbulence is a key for star formation \cite{MacLowKlessen2004,ElmegreenScalo2004,McKeeOstriker2007,Krumholz2014,PadoanEtAl2014}, so most of our simulations include a turbulence driving module that produces turbulence similar to what is observed in real molecular clouds, i.e., driving on the largest scales \cite{HeyerWilliamsBrunt2006,BruntHeyerMacLow2009} and with a power spectrum, $E(k)\sim k^{-2}$, consistent with supersonic, compressible turbulence \cite{Larson1981,HeyerBrunt2004,RomanDuvalEtAl2011} and confirmed by simulations \cite{KritsukEtAl2007,FederrathDuvalKlessenSchmidtMacLow2010,Federrath2013}. We drive turbulence by applying a stochastic Ornstein-Uhlenbeck process \cite{EswaranPope1988,SchmidtHillebrandtNiemeyer2006} to construct an acceleration field ${\bf F_\mathrm{stir}}$, which serves as a momentum and energy source term in the momentum equation of MHD. As suggested by observations, ${\bf F_\mathrm{stir}}$ only contains large-scale modes, $1<\left|\mathbf{k}\right|L/2\pi<3$, where most of the power is injected at the $\kinj=2$ mode in Fourier space, i.e., on half of the box size. The turbulence on smaller scales is not directly affected by the driving and develops self-consistently. The turbulence forcing module used here excites a natural mixture of solenoidal and compressible modes, corresponding to a turbulent driving parameter $b=0.4$ \cite{FederrathDuvalKlessenSchmidtMacLow2010}, although some cloud-to-cloud variations in this parameter from $b\sim1/3$ (purely solenoidal driving) to $b\sim1$ (purely compressive driving) are expected for real clouds \cite{PriceFederrathBrunt2011,KainulainenFederrathHenning2013}.

\begin{table*}
{\small
\caption{Key simulation parameters and measured star formation rates (SFRs).}
\label{tab:sims}
\def\arraystretch{1.1}
\setlength{\tabcolsep}{2.8pt}
\begin{tabular}{lcccccccccc}
\hline
Model Name & Turbulence & $\sigma_v (\km/\s)$ & $\mach$ & $B (\mu\Gauss)$ & $\beta$ & $\macha$ & Jets/Outflows & $N_\mathrm{res}^3$ & $\sfr (\msol/\yr)$ & $\sfrff$ \\
(1) & (2) & (3) & (4) & (5) & (6) & (7) & (8) & (9) & (10) & (11) \\
\hline
G            & None & $0$ & $0$ & $0$ & $\infty$ & $\infty$ & No  & $1024^3$ & $1.6\!\times\!10^{-4}$ & $0.47$ \\
GvsT      & Mix & $1.0$ & $5.0$ & $0$ & $\infty$ & $\infty$ & No  & $1024^3$ & $8.3\!\times\!10^{-5}$ & $0.25$ \\
GvsTM   & Mix & $1.0$ & $5.0$ & $10$ & $0.33$ & $2.0$ & No  & $1024^3$ & $2.8\!\times\!10^{-5}$ & $0.083$ \\
GvsTMJ & Mix & $1.0$ & $5.0$ & $10$ & $0.33$ & $2.0$ & Yes & $1024^3$ & $1.4\!\times\!10^{-5}$ & $0.041$ \\
\hline
GvsTMJ512 & Mix & $1.0$ & $5.0$ & $10$ & $0.33$ & $2.0$ & Yes & $512^3$ & $1.3\!\times\!10^{-5}$ & $0.039$ \\
GvsTMJ256 & Mix & $1.0$ & $5.0$ & $10$ & $0.33$ & $2.0$ & Yes & $256^3$ & $8.9\!\times\!10^{-6}$ & $0.027$ \\
\hline
\vspace{0.0001cm}
\end{tabular}
}
\begin{minipage}{\linewidth}
\textbf{Notes.} Column 1: simulation name. Columns 2--4: the type of turbulence driving \cite{FederrathDuvalKlessenSchmidtMacLow2010}, turbulent velocity dispersion, and turbulent rms sonic Mach number. Columns 5--7: magnetic field strength, the ratio of thermal to magnetic pressure (plasma $\beta$), and the Alfv\'en Mach number. Note that these magnetic quantities are based on the initial uniform magnetic field. However, the total (mean+turbulent) plasma $\beta$ at the end of the simulations is $\beta_\mathrm{end}\sim0.25$ and thus not significantly different from the initial value $\beta=0.33$. This is because the field is relatively strong to begin with, such that turbulent tangling can only amplify the magnetic field by a factor of $(\beta/\beta_\mathrm{end})^{1/2}\sim1.15$ over the course of the whole simulation. Column 8: whether jet and outflow feedback was included or not. Column 9: maximum effective grid resolution (note that refinement is based on the Jeans length with a minimum of 32 cells per Jeans length). Columns 10--11: the absolute SFR and the SFR per mean global freefall time. The standard four simulation models are in the first four rows. Two additional simulations with lower resolution, but otherwise identical to GvsTMJ, are listed in the last two rows, in order to check convergence of our results for the SFR.
\end{minipage}
\end{table*}

\subsection{Sink particles} \label{sec:sinks}
In order to measure the SFR, we use the sink particle method \cite{FederrathBanerjeeClarkKlessen2010}. Sink particles form dynamically in our simulations when a local region in the simulation domain undergoes gravitational collapse and forms stars. This is technically achieved by first flagging each computational cell that exceeds the Jeans resolution density,
\begin{equation}
\rho_\mathrm{sink} = \frac{\pi\cs^2}{G\lambda_\mathrm{J}^2} = \frac{\pi\cs^2}{4 G r_\mathrm{sink}^2},
\end{equation}
with the sound speed $\cs$, the gravitational constant $G$ and the local Jeans length $\lambda_\mathrm{J}$. Thus, the sink particle accretion radius is given by $r_\mathrm{sink} = \lambda_\mathrm{J}/2$ and set to $2.5$ grid cell lengths in order to capture star formation and to avoid artificial fragmentation on the highest level of AMR \cite{TrueloveEtAl1997}. If the gas density in a cell exceeds $\rho_\mathrm{sink}$, a spherical control volume with radius $r_\mathrm{sink}$ is constructed around that cell and it is checked that all the gas within the control volume is Jeans-unstable, gravitationally bound and collapsing towards the central cell. A sink particle is only formed in the central cell of the control volume, if all of these checks are passed. This avoids spurious formation of sink particles and guarantees that only bound and collapsing gas forms stars \cite{FederrathBanerjeeClarkKlessen2010}, which is important for accurately measuring the SFR.

On all the lower levels of AMR (except the highest level, where sink particles form), we use an adaptive grid refinement criterion based on the local Jeans length, such that $\lambda_\mathrm{J}$ is always resolved with at least 32 grid cell lengths in each of the three spatial directions of our 3D domain. This resolution criterion is very conservative and computationally costly, but guarantees that we resolve turbulence on the Jeans scale \cite{FederrathSurSchleicherBanerjeeKlessen2011}, potential dynamo amplification of the magnetic field in the dense cores \cite{SurEtAl2010}, and capture the basic structure of accretion discs forming on the smallest scales \cite{FederrathEtAl2014}. If a cell within the accretion radius of an existing sink particle exceeds $\rho_\mathrm{sink}$ during the further evolution, is bound to the sink particle and is moving towards it, then we accrete the excess mass above $\rho_\mathrm{sink}$ on to the sink particle, conserving mass, momentum and angular momentum.  We compute all contributions to the gravitational interactions between the gas on the grid \cite{Ricker2008} and the sink particles (by direct summation over all sink particles and grid cells). A second-order leapfrog integrator is used to advance the sink particles on a timestep that allows us to resolve close and highly eccentric orbits \cite{FederrathBanerjeeClarkKlessen2010}.

\subsection{Outflow/Jet feedback} \label{sec:jets}
Powerful jets and outflows are launched from the protostellar accretion discs around newborn stars. These outflows carry enough mass, linear and angular momentum to transform the structure of their parent molecular cloud and to potentially control star formation itself. In order to take this most important mechanical feedback effect \cite{KrumholzEtAl2014} into account, we recently extended the sink particle approach such that sink particles can launch fast collimated jets together with a wide-angle, lower-speed outflow component, to reproduce the global features of observed jets and outflows, as well as to be consistent with high-resolution simulations of the jet launching process and with theoretical predictions \cite{FederrathEtAl2014}. The most important feature of our jet/outflow feedback model is that it converges and produces the large-scale effects of jets and outflows already with relatively low resolution, such as with sink particle radii $r_\mathrm{sink}\sim1000\,\AU$ used in our star cluster simulations here. Our module has been carefully tested and compared to previous implementations of jet/outflow feedback \cite{NakamuraLi2007,WangEtAl2010,CunninghamEtAl2011}. The most important difference to any previous implementation is that our feedback model includes angular momentum transfer, reproduces the fast collimated jet component and demonstrated convergence \cite{FederrathEtAl2014}.

\begin{figure*}
\centerline{\includegraphics[width=0.95\linewidth]{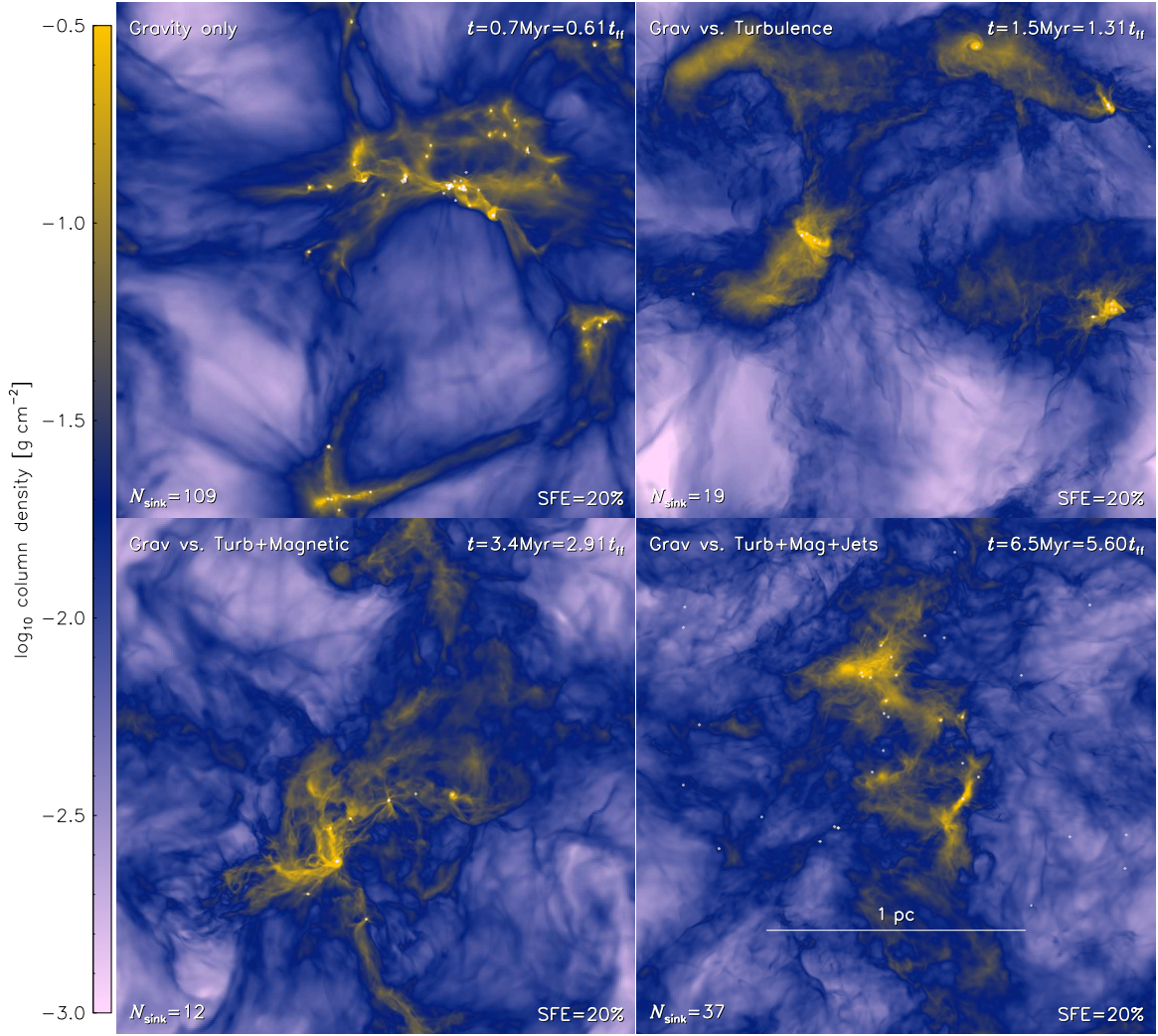}}
\caption{Column density projections at the end of each simulation model: Gravity only (top left), Gravity vs.~Turbulence (top right), Gravity vs.~Turbulence + Magnetic Fields (bottom left), and Gravity vs.~Turbulence + Magnetic Fields + Jet/Outflow Feedback (bottom right). The time to reach a realistic $\sfe=20\%$ with small star clusters having formed, is shown in the top right corner of each panel. The time increases significantly for each model that adds physical processes opposing gravitational collapse. This demonstrates that only the combination of turbulence, magnetic fields and feedback can produces realistic SFRs, which is quantified below and summarized in Table~\ref{tab:sims}. \emph{An animation of these still frames is available at \url{http://astrobites.org/2015/04/28/why-is-star-formation-so-inefficient/}}}
\label{fig:images}
\end{figure*}

\subsection{Simulation parameters}
All our simulations share the same global properties: a cloud size $L=2\,\pc$, a total cloud mass $M=388\,\msol$ and a mean density $\rho_0=3.28\times10^{-21}\,\g\,\cm^{-3}$, resulting in a global mean freefall time $\tff=1.16\,\Myr$. Models including turbulence have a velocity dispersion $\sigma_v=1\,\km\,\s^{-1}$ and an rms Mach number of $\mach=5$, given the sound speed $\cs=0.2\,\km\,\s^{-1}$, appropriate for molecular gas with temperature $T=10\,\mathrm{K}$ over the wide range of densities that lead to dense core and eventually star formation \cite{OmukaiEtAl2005}. Finally, models including a magnetic field start with a uniform initial field of $B=10\,\mu\G$, which is subsequently compressed, tangled and twisted by the turbulence, similar to how it would be structured in real molecular clouds \cite{PlanckMagneticFilaments2015a}. The magnetic field strength, the turbulent velocity dispersion and the mean density all follow typical values derived from observations of clouds with the given physical properties \cite{FalgaronePugetPerault1992,CrutcherEtAl2010}. This leads to the dimensionless virial ratio $\alphavir=1.0$ \cite{KauffmannPillaiGoldsmith2013,HernandezTan2015} and to a plasma beta parameter (ratio of thermal to magnetic pressure) $\beta=0.33$ or an Alfv\'en Mach number $\macha=2.0$. An average Alfv\'en Mach number of about $\macha=1.5$ has been measured in 14 different star-forming regions in the Milky Way \cite{FalgaroneEtAl2008}. Thus, the assumed magnetic field in our simulation models is very close to the values typically observed in molecular clouds and cloud cores.

We run four basic models, which---step by step---include more physics. In the first simulation we only include self-gravity with a given initial density distribution resembling molecular cloud structure, but we do not include any turbulent velocities or magnetic field. In the second model, we include a typical level and mixture of molecular cloud turbulence (see~\S\ref{sec:turbdriving}). The third model is identical to the second model, but adds a standard magnetic field for the given cloud size and mass. Finally, the fourth model is identical to the third model, but additionally includes jet and outflow feedback (see~\S\ref{sec:jets}). These four basic models were all run with a maximum effective grid resolution of $1024^3$ cells. Their key parameters are listed in Table~\ref{tab:sims}. We also run two additional models, which are identical to the fourth model (with jet/outflow feedback), but have a lower maximum effective resolution of $512^3$ and $256^3$ cells, respectively, in order to check numerical convergence of our results for the SFR. Those models are listed in the bottom two rows of Table~\ref{tab:sims}.

% SECTION
\section{Results} \label{sec:results}

\subsection{Cloud structure and stellar distribution}

Figure~\ref{fig:images} shows column density projections of our four basic models from Table~\ref{tab:sims}. The simulation that only includes self-gravity (no turbulence, no magnetic fields and no feedback) is shown in the top left-hand panel. The gas structures we see in the figure resemble the typical distribution found in gravity-only simulations such as simulations of pure dark matter in the cosmological case. Gas from the voids continuously falls on to dense filaments. Stars form along filaments and in particular where filaments intersect. This seems to be true also for real molecular clouds to some degree \cite{SchneiderEtAl2012}, but the inter-filament gas is much more disturbed and turbulent in real molecular clouds compared to the gravity-only case. We stop the simulation when 20\% of the gas is wound up in stars, i.e., the star formation efficiency $\sfe\equiv M_\mathrm{stars}/(M_\mathrm{stars}+M_\mathrm{gas})=20\%$, which is already reached within a fraction of a global mean freefall time, $t=0.61\,\tff$, for the gravity-only simulation. Thus, the whole cloud forms stars in roughly a global freefall time, which is the expected---and unrealistic---outcome if no other physics but gravity is considered.

When we add a realistic level of turbulence to the cloud, shown in the top right-hand panel of Figure~\ref{fig:images}, we find that the time to reach $\sfe=20\%$ increases to $t=1.31\,\tff$ or $1.5\,\Myr$. The overall structure is now much closer to what we see in real molecular clouds, especially the turbulent velocity dispersion measured in observations is approximately reproduced in this model. However, the SFR is still an order of magnitude too large compared to typical observations and there is no magnetic field present, contrary to what is observed \cite{CrutcherEtAl2010}.

Thus, we add magnetic fields in the simulation shown in the bottom left-hand panel of Figure~\ref{fig:images} and see that the end time is shifted by slightly more than a factor of 2, $t=2.91\,\tff=3.4\,\Myr$, compared to non-MHD case. Magnetic fields thus reduce the SFR by about a factor of 2--3, as found in previous simulations \cite{WangEtAl2010,PadoanNordlund2011,PadoanHaugboelleNordlund2012,FederrathKlessen2012}. Although we added a considerable set of physics (gravity, turbulence and magnetic fields) in this simulation model, the SFR is still much higher than inferred from observations, but many spatial and morphological features seen in real star-forming molecular clouds are well reproduced with this set of physics.

Finally, we add jet and outflow feedback in the simulation shown in the bottom right-hand panel of Figure~\ref{fig:images}. The overall cloud structure is similar to the turbulent MHD case without feedback, but the SFR is further reduced by about another factor of 2. Now it takes $6.5\,\tff=5.6\,\Myr$ to reach a reasonable $\sfe=20\%$ in this small cluster-forming region of a molecular cloud. The most interesting morphological difference compared to all the no-feedback cases is that the oldest stars have wandered away from their formation sites and are now distributed over a much wider volume. Newborn stars are still always located on or very close to high-density peaks. A total of 37 stars have formed in the whole field, while some of those stars might actually represent close binaries or small multiple systems given the limited resolution available in these calculations. While the absolute fragmentation is not fully converged in these simulations, we emphasize that the SFR and the amount of mechanical feedback is well captured and is converged with our numerical resolution. This is shown in Table~\ref{tab:sims}, where in the last two rows, we compare this feedback model (GvsTMJ) with two lower-resolution equivalents (GvsTMJ512 and GvsTMJ256), demonstrating convergence of the SFR to within $5\%$.

\subsection{Star formation rate}

\begin{figure}
\centerline{\includegraphics[width=1.0\linewidth]{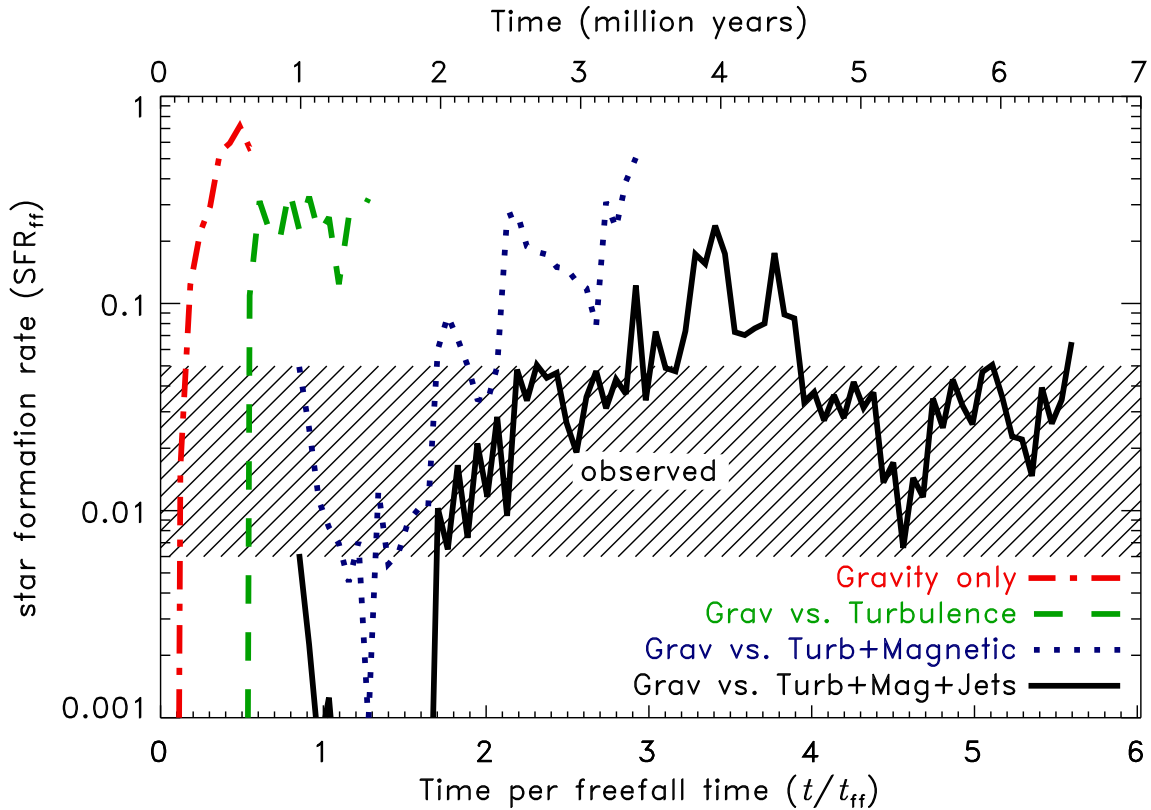}}
\caption{Time evolution of the SFR per freefall time, $\sfrff=d(\sfe)/d(t/\tff)$ for each of our four standard models from Table~\ref{tab:sims}. We immediately see the drastic reduction in the SFR when turbulence, magnetic fields and jet/outflow feedback together counteract gravity, in which case the SFR is reduced to values consistent with the observed range of SFRs, shown as the hatched region.}
\label{fig:tevol}
\end{figure}

Now we concentrate on the quantitative determination of the SFR. Figure~\ref{fig:tevol} shows the time evolution of the $\sfr$ in our four fiducial simulation models. While the gravity-only model (dot-dashed line) turns all the gas into stars in about a global mean freefall time, the SFR per freefall time defined as $\sfrff\equiv d(\sfe)/d(t/\tff)$ \cite{KrumholzMcKee2005,FederrathKlessen2012} is reduced by about a factor of 2 when turbulence is included (dashed line). The $\sfrff$ is reduced by at least another factor of 2 when turbulence and magnetic fields are present (dotted line), and by another factor of 2 when jet and outflow feedback is included (solid line).

The hatched region in Figure~\ref{fig:tevol} shows the observed range of $\sfrff$ based on the observational data compiled in Krumholz \& Tan (2007) \cite{KrumholzTan2007}. Our feedback model eventually settles into that observed range.

A very interesting and noteworthy feature of our feedback model is the evolution of the SFR with time, in particular the fact that the SFR slowly but steadily increases until about $t\sim4.0\,\Myr$ and then drops significantly by almost an order of magnitude until $t\sim5.3\,\Myr$, followed by a period of nearly constant SFR. This is a manifestation of self-regulation by feedback. As more material is accreted and the SFR increases, the amount of gas being re-injected into the interstellar medium in the form of fast mass-loaded jets and outflows increases proportionally. Eventually, the feedback enhances the turbulence and the amount of kinetic energy in the system, thereby increasing the virial parameter, $\alphavir=2E_\mathrm{kin}/E_\mathrm{grav}$, such that star formation is rapidly quenched. This quenching of the SFR in turn reduces the amount of feedback such that the system eventually settles into a self-regulated state of star formation in which any intermittent increase in accretion triggers feedback that regulates the SFR down to a nearly constant level.

The last two columns in Table~\ref{tab:sims} list the time-averaged SFRs in each simulation model. For the gravity-only model, we find unrealistically high $\sfrff=0.47$, while the feedback model including turbulence and magnetic fields has a time-averaged $\sfrff=0.041$. Thus,cd the combination of turbulence, magnetic fields, and jet/outflow feedback reduces the SFR by more than an order of magnitude and brings the SFR into the observed range.

% SECTION
\section{Summary and conclusions} \label{sec:conclusions}

We presented high-resolution hydrodynamical simulations of star cluster formation including turbulence, magnetic fields and jet/outflow feedback. Although our simulation models with turbulence, magnetic fields and jet/outflow feedback produce $\sfrff\sim0.04^{+0.04}_{-0.02}$, which is currently the closest available match to observations, they still overproduce stars by a factor of 2--4 compared to the observed average $\sfrff\sim0.01$. This discrepancy might be resolved by considering other types of feedback in addition to jet/outflow feedback, such as radiation pressure, which seems to be capable of reducing the SFR further by another factor of two \cite{MacLachlanEtAl2015}. We conclude that only the combination of strong turbulence with virial parameters above unity, strong magnetic fields and mechanical plus radiative feedback can produce realistic SFRs consistent with observations.

% ACKNOWLEDGEMENTS
\section*{Acknowledgements}
\vspace{0.1cm}
C.F.~acknowledges funding provided by the Australian Research Council's Discovery Projects (grants~DP130102078 and~DP150104329).
The author gratefully acknowledges the J\"ulich Supercomputing Centre (grant hhd20), the Leibniz Rechenzentrum and the Gauss Centre for Supercomputing (grants~pr32lo, pr48pi and GCS Large-scale project~10391), the Partnership for Advanced Computing in Europe (PRACE grant pr89mu), and the National Computational Infrastructure (grant ek9), supported by the Australian Government. This work was further supported by resources provided by the Pawsey Supercomputing Centre with funding from the Australian Government and the Government of Western Australia. The software used in this work was in part developed by the DOE-supported Flash Center for Computational Science at the University of Chicago.

\section*{References}
\vspace{0.1cm}

%\bibliographystyle{iop}
%\bibliographystyle{/Users/chfeder/Documents/Latex-sources/apj}
%\bibliography{/Users/chfeder/Documents/Latex-sources/federrath.bib}

\end{document}